\documentclass[reprint,amsmath,amssymb,twocolumn,superscriptaddress]{revtex4-1}

\usepackage[pdftex]{graphicx}
\usepackage{siunitx}

\begin{document}

\title{Probing electron acceleration and X-ray emission in laser-plasma accelerator}

\author{C. Thaury}
\affiliation{Laboratoire d'Optique Appliqu\'ee, ENSTA ParisTech - CNRS UMR7639 - \'Ecole Polytechnique ParisTech, Chemin de la Huni\`ere, 91761 Palaiseau, France}
\author{K. Ta Phuoc}
\affiliation{Laboratoire d'Optique Appliqu\'ee, ENSTA ParisTech - CNRS UMR7639 - \'Ecole Polytechnique ParisTech, Chemin de la Huni\`ere, 91761 Palaiseau, France}
\author{ S. Corde}
\affiliation{Laboratoire d'Optique Appliqu\'ee, ENSTA ParisTech - CNRS UMR7639 - \'Ecole Polytechnique ParisTech, Chemin de la Huni\`ere, 91761 Palaiseau, France}
\author{ P. Brijesh}
\affiliation{Laboratoire d'Optique Appliqu\'ee, ENSTA ParisTech - CNRS UMR7639 - \'Ecole Polytechnique ParisTech, Chemin de la Huni\`ere, 91761 Palaiseau, France}
\author{G. Lambert}
\affiliation{Laboratoire d'Optique Appliqu\'ee, ENSTA ParisTech - CNRS UMR7639 - \'Ecole Polytechnique ParisTech, Chemin de la Huni\`ere, 91761 Palaiseau, France}
\author{S.P.D. Mangles}
\affiliation{Blackett Laboratory, Imperial College, London SW7 2AZ, United Kingdom}
\author{M. S. Bloom}
\affiliation{Blackett Laboratory, Imperial College, London SW7 2AZ, United Kingdom}
\author{S. Kneip}
\affiliation{Blackett Laboratory, Imperial College, London SW7 2AZ, United Kingdom}

\author{ V. Malka}
\affiliation{Laboratoire d'Optique Appliqu\'ee, ENSTA ParisTech - CNRS UMR7639 - \'Ecole Polytechnique ParisTech, Chemin de la Huni\`ere, 91761 Palaiseau, France}

\begin{abstract}
While laser-plasma accelerators have demonstrated a strong potential in the acceleration of electrons up to giga-electronvolt energies, few experimental tools for studying the acceleration physics  have been developed. 
In this paper, we demonstrate a method for probing  the acceleration process. A second laser beam, propagating perpendicular to the main beam is focused in the gas jet few nanosecond before the main beam   creates the accelerating plasma wave. This second beam is intense enough to ionize the gas and form a density depletion which will locally inhibit the acceleration. The position of the density depletion is scanned along the interaction length to probe the electron injection and acceleration, and the betatron X-ray emission.  To illustrate the potential of the method, the variation of the injection position with the plasma density is studied.
\end{abstract}

\maketitle

Over the past decade,  considerable progress has been made in the development of laser-plasma accelerators~\cite{RMP2009Esarey}. These accelerators can now deliver quasi-mono energetic electron beams~\cite{Nature2004Faure,*Nature2004Mangles,*Nature2004Geddes} with energy up to 1 GeV~\cite{NatPhys2006Leemans} and energy spread close to 1\%~\cite{PRL2009Rechatin1}. A significant step forward has also been  achieved in the control and the stabilization of the accelerator using controlled injection techniques (e.g. colliding pulses injection~\cite{Nature2006Faure}, down-ramp injection~\cite{PRSTAB2011Schimdt,NatPhys2011Gonsalves}, or ionization injection~\cite{PRL2011Pollock} ).  However more improvements, in particular in terms of stability and  energy spread,  are needed  for laser-plasma accelerators to compete with state of the art conventional accelerators.
To further push  the performance of laser-plasma accelerators and enhance control over the interaction, a thorough knowledge 
of the details of the interaction is required.

In a laser plasma accelerator, electrons are injected in a relativistic plasma wave which is excited in the wake of an intense laser pulse propagating in an underdense plasma. The electric field associated with the plasma wave has a very large amplitude, of a few hundreds of gigavolt per meter. Electrons can therefore be accelerated to a few hundreds of~MeV within just a few millimeters. Three phenomena play an important role in a laser plasma accelerator: the laser propagation; the electron injection; and the electron acceleration itself. All these phenomena are difficult to experimentally characterize, which can be largely traced back to the very small size of the accelerator and to the extremely fast (femtosecond) dynamics of the interaction.  
  As a result, current knowledge of the laser-plasma interaction  relies heavily on numerical simulations. This lack of experimental information could hinder the development of laser-plasma accelerators. New experimental tools were therefore recently developed to get more insight into the acceleration process. 
 Among these, the observation of the shadow of a small aperture illuminated with Betatron X-ray radiation was used to map the acceleration length and injection position~\cite{genoud2011,2011prl_cordeA}. The acceleration phase was also probed by varying the acceleration length, using different methods~\cite{Nature2006Faure,IEEE2008Faure,PRL2006Hsieh,PRLKneip2009}. Further information about the plasma wave was obtained from interferometric~\cite{matlis2006np} and	magnetic fields~\cite{PRL2010kaluza,2011NatPh_buck} measurements. Techniques were also developed to determine whether electrons are trapped in the first plasma wave period~\cite{PRL2006Mangles}, whether they are trapped in two or more periods~\cite{PRL2007Glinec} and to determine the position of injection~\cite{PRL2007Thomasa}.
In this letter, we present a method for probing the acceleration. It relies on the use of a second laser beam to locally disrupt  the accelerating structure and hence assess the influence of a given part of the plasma on electron acceleration and X-ray emission. 

\begin{figure}
	\centering
		\includegraphics[width=\linewidth]{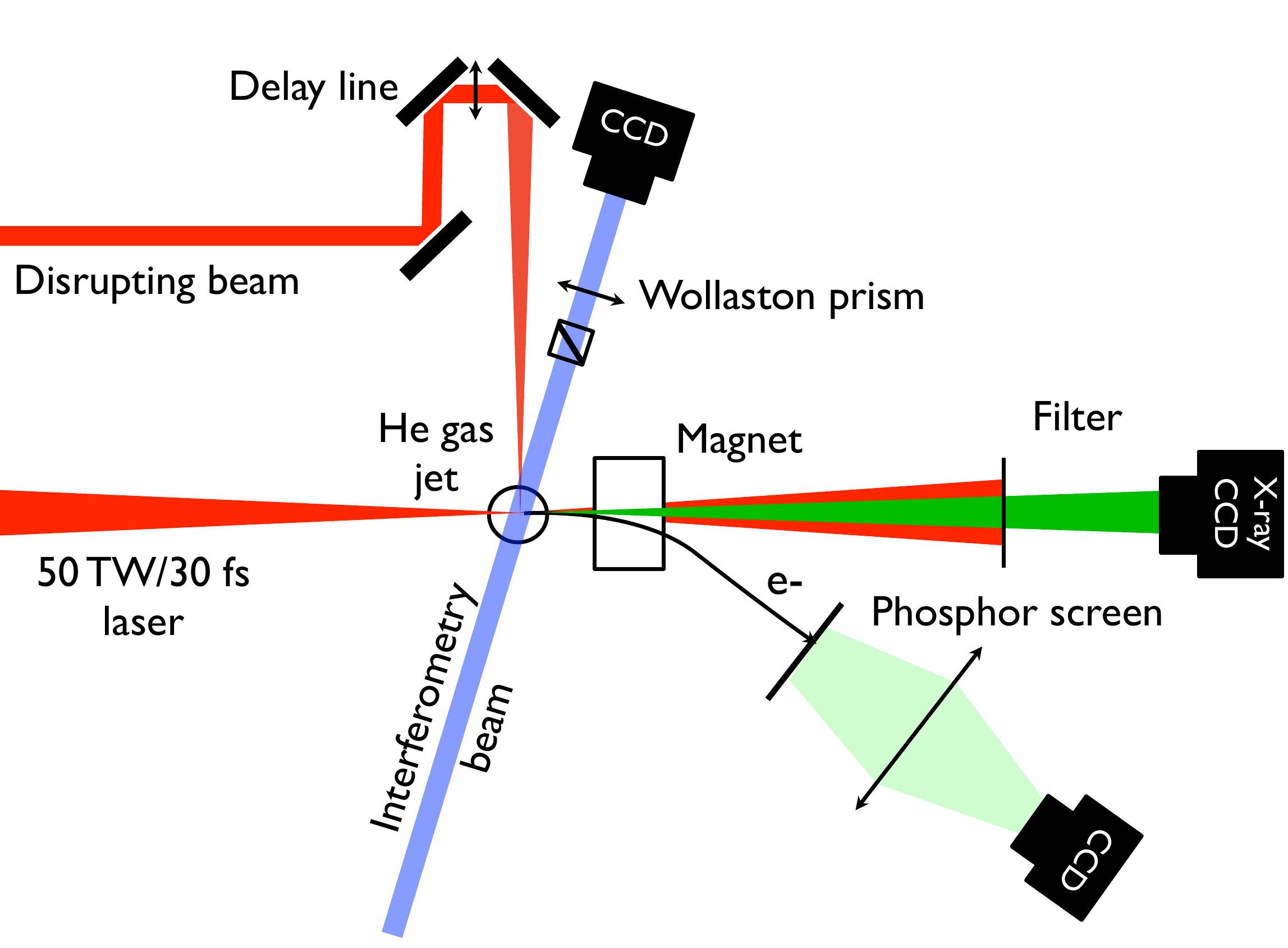}
	\caption{(Color online) Experimental setup.} 
	\label{fig:fig1}
\end{figure}
The experiment was performed at Laboratoire d'Optique Appliqu\'ee using the ``Salle Jaune'' Ti:Sa laser system, which delivers two laser pulses (main and disrupting beams) with a full width at half maximum (FWHM) duration of 35 fs and a linear polarization.  The main laser pulse had an energy of 900 mJ. It was focused by a $F/15$ spherical mirror in a 3~mm supersonic helium gas jet. The FWHM focal spot size was $14 \times 18$  \si{\micro\meter}$^2$, and the peak intensity was about $4.6\times10^{18}\:\textrm{W.cm}^{-2}$, corresponding to a normalized amplitude $a_0=1.5$. The disrupting pulse had an energy of 100 mJ. It made an angle of 90$^\circ$ with the main beam (see Fig.~\ref{fig:fig1}) and was focused into the gas jet with a $F/10$ plano-convex lens to an intensity of $\approx 7\times10^{17}\:\textrm{W.cm}^{-2}$. The intensity of the disrupting laser pulse was thus much larger than in experiments where the plasma channel created by a second pulse is used to trigger the electron injection~\cite{pop_faure_2010,pop_brijesh} (the electron density in the present experiment is also larger). The disrupting beam entered the plasma 2 ns before the main beam. Electron beam spectra were measured with a spectrometer consisting of a dipole magnet (1.1 T over 10 cm) and a LANEX phosphor screen. 
Betatron X-ray beam profiles were measured using a  X-ray CCD camera with $2048\times2048$ pixels of size 13.5  \si{\micro\meter} $\times$ 13.5 \si{\micro\meter}, situated 73 cm from the gaz jet and protected from the laser light by a 13 \si{\micro\meter} Al filter. 
The density depletion created by the disrupting beam was observed using a Wollaston interferometer.

\begin{figure}
	\centering
		\includegraphics[width=\linewidth]{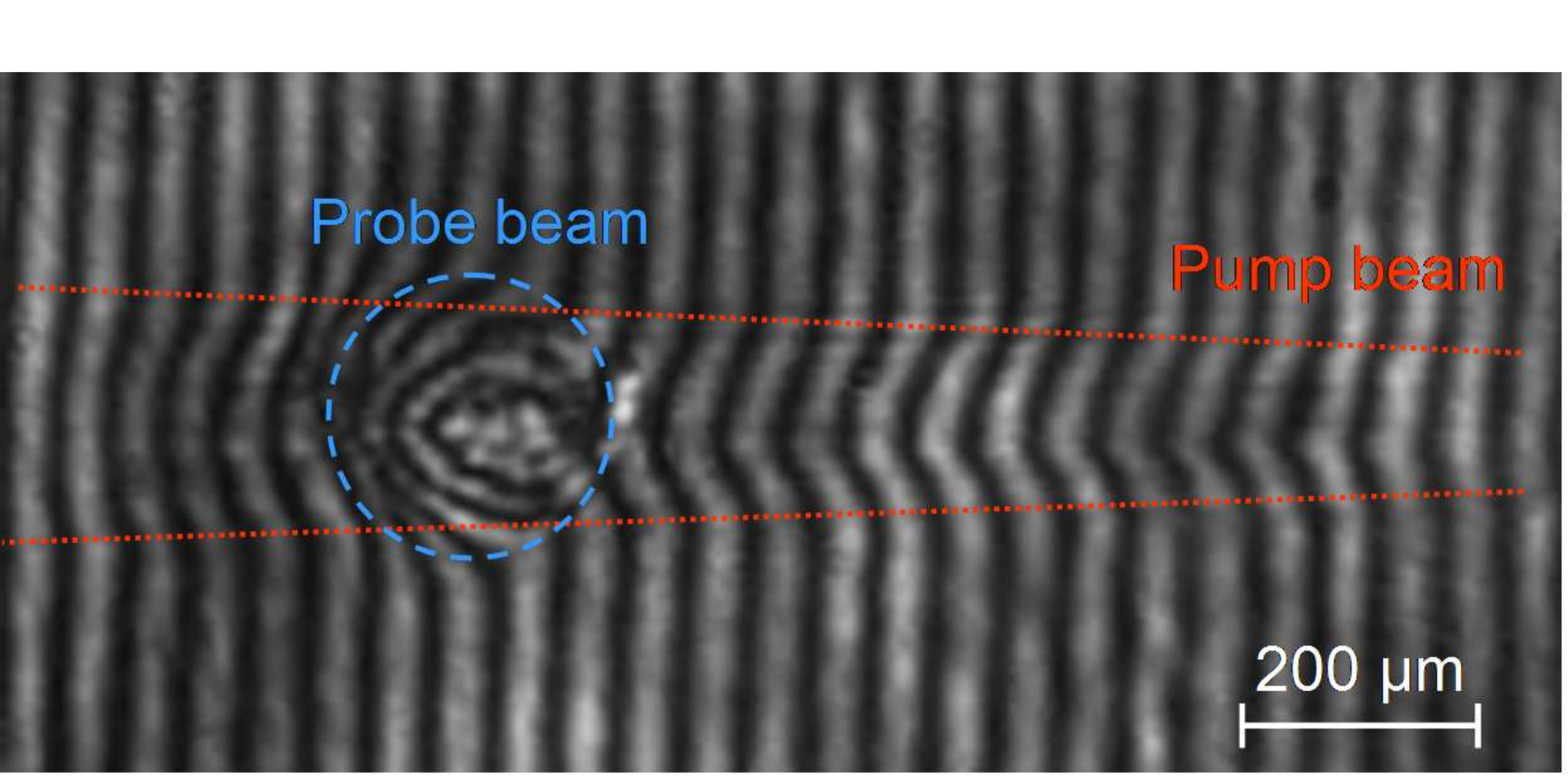}
	\caption{(Color online) Interferometric image of the plasma obtained $\approx 5.6$ ps after the main laser pulse has entered the gas jet. The laser goes from right to left. The dotted red lines highlight the beam path. The width of the density depletion zone created by the disrupting beam (dashed blue circle) is  $\delta =255 \pm 15$ \si{\micro\meter}.} 
	\label{fig:fig2}
\end{figure}

Figure~\ref{fig:fig2} shows an interferogram of the plasma obtained a few picoseconds after the main beam has entered the gas jet. The plasma formed by the main beam is visible between the two dotted lines. The disrupting beam propagates perpendicular to the main beam and ionizes the helium gas along its path. With time, this
plasma expands hydrodynamically, which creates a density depression at the center of the plasma column.  This depletion zone is clearly observed in Fig.~\ref{fig:fig2} (dotted circle). The full width of this zone is  $\delta \approx 250$ \si{\micro\meter}. The amplitude of the depletion cannot be determined precisely because the phase shift is too large. We will show in the following that it is however sufficient to prevent electron injection and acceleration locally, in contrast with Refs.~\cite{pop_faure_2010,pop_brijesh}. The electron acceleration process can thus be probed by moving the position of the disrupting beam along the main beam path to inhibit the acceleration in a given zone of the plasma and hence assess the influence of this zone on the acceleration. 

\begin{figure}
	\centering
		\includegraphics[width=\linewidth]{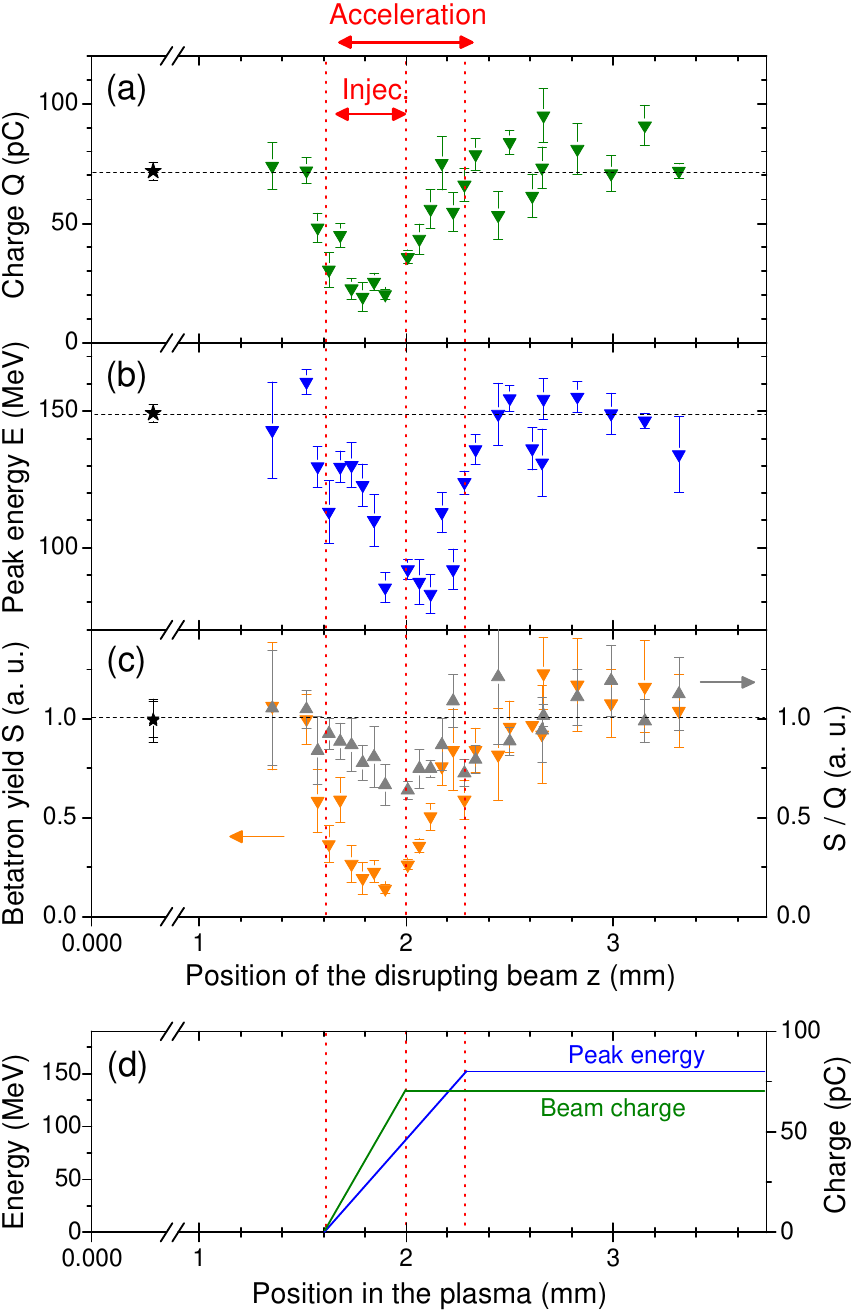}
	\caption{ Beam charge $Q$ (a), peak electron energy $E$ (b), betatron yield $S$ and quantity $S/Q$ (c) as a function of the position in the plasma for $n_e=10^{19}$ cm$^{-3}$. The position $z=0$ corresponds to the entrance of the gas jet.  The red dashed  lines indicate the estimated injection and acceleration lengths. The blacks stars correspond to shots without the disrupting beam. The black dashed lines indicate the mean charge and energy without disrupting beam. Each point corresponds to an average over 5 to 27 shots. The error bars correspond to standard errors of the mean. (d) Schematic of possible injection and acceleration profiles.}
	\label{fig:fig3}
\end{figure}

Figure~\ref{fig:fig3} shows the beam charge $Q$ and the peak electron energy $E$ as a function of the disrupting beam position $z$, for an electron density $n_e=10^{19}$ cm$^{-3}$. In this experiment electron spectra were broad (because of a long injection length due to relatively high densities). 
For $z\lesssim1.6$~mm, the disrupting beam has no influence on the electron beam features ($E$ and $Q$ are the same as without disrupting beam). This indicates that no electrons are injected in this part of the plasma. The intensity of the main laser pulse for $z\lesssim 1.6$~mm is not sufficient to trigger the injection of electrons into the relativistic plasma wave. As the main laser pulse propagates in the plasma, it self-focuses and eventually reaches an intensity high enough to trigger the injection for $z\gtrsim1.6$~mm. For $1.6$~mm $\lesssim z\lesssim2$~mm, $Q$ is strongly reduced, which demonstrates that the disrupting beam inhibits locally the plasma wave and prevents electron injection in the depletion zone.  From $z\approx 2$~mm, $Q$ increases until it reaches its initial value for $z\gtrsim 2.2$~mm. The influence of the disrupting beam on the injection is thus less dramatic in this part of the plasma. An estimate of the injection length $ L_{inj}$ can be obtained from these measurements. The function $Q(z)$ is actually a convolution of the injection zone with the depletion profile. Assuming that the depletion profile can be described by a square function of width $\delta$, we find that $L_{inj}\approx 0.4$~mm. Thus the disrupting beam reduces the injection length by $\delta/L_{inj} \approx 70\%$, which  is consistent with a decrease of $Q$ from  $72$ pC down to $21$ pC at minimum.

The evolution of the peak electron energy $E$ in Fig.~\ref{fig:fig3}b is somewhat different from that of the electron charge $Q$. The energy remains small for a longer length than $Q$, and it comes back to its initial value only for $z\gtrsim 2.4$~mm (instead of $\approx 2$~mm for $Q$). For $2$~mm $\gtrsim z\gtrsim 2.4$~mm, very few electrons are injected but electrons which have been injected earlier are still accelerated in the plasma wave. Focusing the disrupting beam in this region shortens the total acceleration and hence reduces the peak electron energy $E$. For $z\gtrsim 2.4$~mm, the disrupting beam has no effect on $E$, which suggests that the laser intensity has decreased down to a level at which it can no more sustain a plasma wave, so that the acceleration stops. The acceleration length $L_{acc}$ is estimated to be $L_{acc}\approx 0.6$~mm. The shortening of the acceleration length due to the disrupting beam is $\delta/L_{acc} \approx 40\%$, in good agreement with the decrease of $E$ from $150$~MeV down to $87$~MeV. A  schematic of possible injection and acceleration profiles that would be consistent with the experimental measurements is displayed  in Fig.~\ref{fig:fig3}d.

Also shown in Fig.~\ref{fig:fig3} is  the X-ray emission yield (energy in the beam integrated over angles and frequencies) due to betatron transverse oscillations of  relativistic electrons into the plasma wave~\cite{PRL2004Rousse}. 
The variations of the X-ray yield $S$  (left axis in Fig.~\ref{fig:fig3}c)  and of the electron charge $Q$ (Fig.~\ref{fig:fig3}a) are very similar, which indicates that $S$ depends mainly on the electron charge. Some influence of the electron peak energy $E$ is however noticed for $z\gtrsim 1.9$ mm. To assess more quantitatively the influence of $E$ on $S$ we plotted in Fig.~\ref{fig:fig3}c (right axis) the betatron yield by unit of charge $S/Q$. As anticipated, the quantity $S/Q(z)$ is observed to vary approximately as $E(z)$ (Fig.~\ref{fig:fig3}b). However, the dependence of $S$ on $E$ is much weaker than reported in previous works~\cite{2011prl_cordeB}. The betatron yield is divided only by $1.5$ when the electron energy is decreased from $150$ MeV down to $85$ MeV, while in Ref.~\cite{2011prl_cordeB} the betatron yield around $3$ keV is divided by more than $10$ for the same variation of $E$. Three main reasons can explain this difference.  First, the interaction regime is different. The the electron spectrum is broad while in Ref.~\cite{2011prl_cordeB} quasi-mono-energetic electron beams were accelerated. In Fig.~\ref{fig:fig3}c the variation of the betatron yield with the electron energy is thus convolved with the electron spectrum, which tends to smooth the variations of $S$ with $E$. Second, $S$ in Fig.~\ref{fig:fig3}(c) is integrated over a  large spectral bandwidth while in Ref.~\cite{2011prl_cordeB} a narrow spectral window is selected.  The betatron critical energy and hence the whole betatron spectrum shifts to higher frequencies when $E$ increases. As the quantum efficiency of the X-ray camera decreases for $E\gtrsim 3$ keV, our detection system tends to underestimate the increase of the betatron yield with the electron energy.Lastly, the amplitude of the betatron oscillations  is increased in the density depletion, which can enhance the betatron emission~\cite{2008PhPl...15f3102T}. This effect may counterbalance the decrease of $S$ due to the shortening of the acceleration length and thus may contribute to explain the weak dependence of $S$ on $E$. To estimate the significance  of this last effect and demonstrate an enhancement of the betatron emission, the ratio $\delta/L_{acc}$ should be decreased.  More generally the strong similarity between the variations of $S$, $Q$ and $E$ confirms the strong correlation between relativistic electrons and  betatron X-rays~\cite{2011prl_cordeB}. It shows also that the measure of X-ray emission can provide  information on the acceleration.

\begin{figure}
	\centering
		\includegraphics[width=\linewidth]{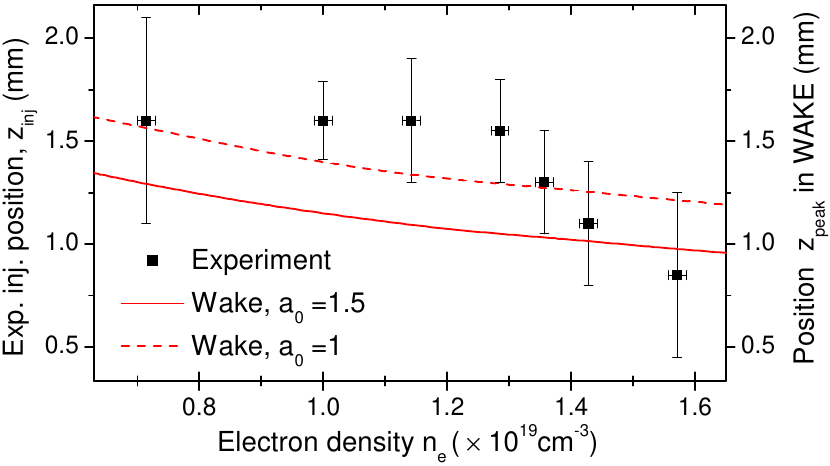}
	\caption{Injection position $z_{inj}$ and peak laser intensity position $z_{peak}$ as a function of the plasma density. The squares are experimental injection positions. The red lines correspond to the position of the peak laser intensity obtained from {\sc wake} simulations performed for $a_0=1.5$ and $a_0=1$. }
	\label{fig:fig4}
\end{figure}

We observed in Fig.~\ref{fig:fig3}	 that for an electron density $n_e=10^{19}$ cm$^{-3}$, the injection begins at $z\approx 1.6$~mm from the gaz jet entrance. In other words $1.6$~mm are required for the laser to reach an intensity sufficient to trigger the injection. As the laser propagation is largely determined by self-focusing and self-compression which both depend on the plasma density $n_e$, the injection position $z_{inj}$ is also expected to vary with $n_e$. To check this $z_{inj}(n_e)$ is plotted in Fig.~\ref{fig:fig4} for $7\times 10^{18}$ cm$^{-3} <n_e<1.6\times 10^{19}$ cm$^{-3}$. These  data are compared with the position of the peak laser intensity $z_{peak}$ obtained from  {\sc wake} simulations~\cite{pop97mora}. {\sc wake} is a two-dimensional cylindrical relativistic particle code. It describes the interaction of  a laser pulse with an underdense plasma using the quasi-static approximation. The laser pulse is gaussian in time and space with a FWHM duration of 35 fs and a focal spot size of 16 \si{\micro\meter} FWHM. The measured density profile is modeled by a trapezoid with 700 \si{\micro\meter} ramps and a full length of 3.5~mm. The laser is focused 500 \si{\micro\meter} beyond the gas jet center.

 As expected $z_{inj}$ and $z_{peak}$ both decrease when $n_e$ increases.  However, for $a_0=1.5$ the position $z_{peak}$ is significantly smaller than $z_{inj}$. To get a better agreement between $z_{inj}$ and $z_{peak}$  smaller $a_0$ have to be used. Figure~\ref{fig:fig4} also suggests that $z_{inj}$ saturates at low density. This effect is not observed in simulations for the considered parameters.  Such discrepancies between experiment and simulations were also observed in previous works~\cite{2011prl_cordeA} where experimental data were compared to three dimensional Particle-In-Cell simulations. They may be due to an  imperfect modeling of the experimental conditions (\emph{e.g.} the modeling of the laser pulse by a Gaussian)~\cite{2012PPCFVieira}, and to fact that WAKE does not describe the complex injection dynamics~\cite{kalmykov2009PRL}. Nevertheless, these discrepancies demonstrate a need for innovative tools to study laser plasma acceleration. An accurate knowledge of the dynamics of injection is clearly required before  control of the injection, in particular in the wavebreaking regime.

In conclusion we demonstrated a method for probing the electron injection and acceleration lengths, as well as X-ray emission, in laser-plasma accelerators.
This probing technique is similar to the one described in Ref.~\cite{PRL2006Hsieh} where a second laser beam,  focused on a line, is used to modify the acceleration length. Our technique presents several advantages. First, a local disruption provides information that cannot be obtained by changing the acceleration length. In particular, the method described in Ref.~\cite{PRL2006Hsieh} does not allow to measure the injection profile, because for very short plasma lengths the electron energy is too low for electrons to be detected. With the method proposed here,  only the part of the plasma where electrons are injected is disrupted. Electrons are still accelerated on a substantial length, allowing to probe the injection region.  Second, the disrupting laser beam can be much less energetic than the machining beam in Ref.~\cite{PRL2006Hsieh}, because it is focused on a point rather than on a line. The setup is also simpler and easier to align. For instance, in Ref.~\cite{PRL2006Hsieh} the line focus and the main beam path have to be perfectly superimposed, while here the position of the disrupting beam can be easily changed to probe a different region. The main advantage of the line focus setup is to allow the creation of structured plasmas  for optimizing the acceleration~\cite{hung2012}.

Compared to the method presented in Ref. \cite{2011prl_cordeA}, our new technique has the disadvantages that it is multi-shot,  requires a second beam and  modifies the interaction. It has however several advantages. The main one is that it provides direct and separate information on injection and acceleration, without using any model. Moreover the earlier method~\cite{2011prl_cordeA} needs a small aperture mask to be placed close to the accelerator exit. This can be difficult to achieve in many cases. Further the aperture mask erodes a bit after each shot,  which hinders the precision of the measure. The new method is free of such limitations. Lastly the limit on  resolution of the new method is likely to be less stringent. To improve this resolution, different disrupting beam features (intensity, spot size, duration) and delays between the disrupting and the main beams should be tested.

The research leading to these results has received funding from the European Research Council (PARIS ERC project) and LASERLAB-EUROPE (grant agreement n° 228334, EC's Seventh Framework Program).


%

\end{document}